# Systematic Solutions to Login and Authentication Security Problems: A Dual-Password Login-Authentication Mechanism

Suyun Borjigin
Independent Researcher, China
yunsu000@hotmail.com

**Abstract---** Credential theft and remote attacks are the most serious threats to user authentication mechanisms. The crux of these problems is that we cannot control such behaviors. However, if a password does not contain user secrets, stealing it is useless. If unauthorized inputs are invalidated, remote attacks can be disabled. Thus, credential secrets and account input fields can be controlled.

Rather than encrypting passwords, we design a dual-password login-authentication mechanism, where a user-selected secret-free login password is converted into an untypable authentication password. Subsequently, the authenticatable functionality of the login password and the typable functionality of the authentication password can be disabled or invalidated to prevent credential theft and remote attacks. Thus, the usability-security tradeoff and password reuse issues are resolved; local authentication password storage is no longer necessary.

More importantly, the password converter acts as an open hashing algorithm, meaning that its intermediate elements can be used to define a truly unique identity for the login process to implement a novel dual-identity authentication scheme. In particular, the system-managed elements are concealed, inaccessible, and independent of any personal information and therefore can be used to define a perfect unforgeable process identifier to identify unauthorized inputs.

## 1 INTRODUCTION

Since the advent of user authentication technology, the technical architecture of the "one password plus login and authentication" process has not changed. However, the exclusive use of passwords has suffered from the usability-security tradeoff dilemma [1]. The main problem with single-password authentication mechanisms is that most security threats [2, 3, 9] are largely derived from this dilemma. Despite the ongoing efforts [4-6, 11] of the cybersecurity community, this issue is still unsolved [7].

Passwords have dominated client login and server authentication schemes for more than half a century. The usability-security tradeoff and password encryption algorithms [2, 8] have received considerable research attention. However, decades of practice have shown that striking a balance between usability and security and encrypting only passwords are not sufficient for guaranteeing the security of a single-password authentication mechanism.

We believe that the tradeoff is systematic in nature, and all the related problems can be attributed to one issue, i.e., the use of one password in an attempt to simultaneously implement its typable and authenticatable functionalities. As the typable functionality is characterized by usability and the authenticatable functionality is described by security, these functionalities also come at the expense of one another. Provided that individual passwords continue to suffer from this tradeoff, their functionalities will suffer as well. Unfortunately, it is nearly impossible to solve these problems with a single-password authentication mechanism. As noted above, an authentication mechanism involves two processes, which seems to imply that a feasible solution for solving the tradeoff might be the use of one more password and its corresponding identity in a novel dual-identity authentication mechanism.

However, little attention has been given to the fact that in any user authentication mechanism, each of the login and authentication processes requires only one password functionality, either the typable or authenticatable functionality. The login process only needs a user to enter their password (i.e., a login password) to log in rather than authenticating this password, meaning that only the typable functionality is required during the login process. In contrast, the authentication process requires only the authenticatable functionality of a password (i.e., an authentication password) to authenticate the user instead of logging them in. In other words, the authenticatable functionality is not required for the login password, and the typable functionality is not necessary for the authentication password. However, these unnecessary functionalities are essential for implementing credential theft and remote attacks [2, 3, 9, 10, 28]. Unfortunately, in a single-password user authentication mechanism, they are available for attackers.

Rather than encrypting textual passwords, we establish a dual-password login-authentication mechanism by integrating a pair of login and authentication passwords created based on a single-character conversion technique [11]. Utilizing this technique, a set of login characters (i.e., a login password) selected by a user can be randomized twice and converted into an authentication password. After integrating the system with both passwords, the login password can be configured to perform only the login process, whereas the authentication password can be configured to perform only the authentication process, which runs within the isolated environment of the system and is initiated by the login process. Consequently, the login process no longer needs the authenticatable functionality of the login password, whereas the authentication process no longer requires the typable functionality of the authentication password. In this paper, we aim to utilize these unnecessary functionalities to prevent credential theft and remote attacks.

After the creation of the password pair, the two passwords can be further set to be completely different from each other in terms of password strength [2, 12, 13, 19-22] based on the characteristics of the single-character conversion process. Subsequently, the password field of the user interface can be configured to only accept the login password, which can be defined to contain lowercase letters and/or digits, and to reject the authentication password, which can be defined to contain at least four character classes (such as uppercase letters, lowercase letters, digits and symbols) [12, 16, 19, 20]. That is, in the proposed dual-password login-authentication mechanism, any characters other than lowercase letters and digits are invalid, and inputs including such invalid characters are invalidated by the password field. Thus, the highest levels of login usability and authentication security can be simultaneously achieved for the password pair.

Upon integrating the system with the pair of passwords, the unnecessary functionalities are actually discarded. Given their indispensability for carrying out credential theft and remote attacks, they cannot be left unchecked. Rather, they must be either disabled or invalidated. To this end, the login password is not associated with the publicly available identity of the user, such as their name, email address, phone number, etc., making it a secret-free password or unauthenticatable. Moreover, any authentication password inputs can be invalidated by preventing the password field from accepting the invalid characters, making it an untypable password. The result of such a configuration is that the pair of passwords is useless to malicious parties, and therefore, the highest levels of secure login usability and usable authentication security are simultaneously achieved for the client and the server.

More importantly, as the dual-password mechanism involves two passwords, it is possible to define different identities for them. Thus, the identity of the user may be associated with the authentication password, which is related to granting the user access to the password page and later to the user account at the system server. In contrast, a novel login process identity may be associated with the login password, which is related to giving the login process permission to initiate the authentication process.

In particular, the process identifier is concealed, inaccessible, and independent of any personal information that is readily available to the public since it can be selected only by the system from the intermediate elements of a quasi-matrix password converter (Section 7.2). Therefore, the use of a process identifier with the above properties can guarantee the security of initiating the authentication process, and it is highly unlikely that the process identifier can be duplicated or forged by anyone.

When a user interacts with the dual-password login-authentication system using their login credentials, a system routine is called to check the system database for a match of their login credentials. When a username or password page mismatch is detected, it can be identified as a nonlocal login attempt, and the system then initiates the corresponding security measure to invalidate it. For example, in response to an input in the username field, if a mismatch of the smartphone identifier of the user is detected by the routine, the password field will be grayed out when reaching the password page to disable any further inputs, thereby invalidating the identified nonlocal login attempt. In response to an input on the password page, if a mismatch of the process identifier is detected, the identified nonlocal login attempt will be locked out of initiating the authentication process. Furthermore, the system routine can also detect a password strength violation on the password page so that the identified nonlocal login attempt can be prevented from initiating the authentication process. Thus, it is impossible to log in using any passwords during nonlocal login attempts.

The dual-password login-authentication system is designed mainly to perform user registration and login on non-prepaid smartphones [14], while other devices can only be logged into the system by the registered smartphone. This work aims to address the following issues: the usability-security tradeoffs of singe-password authentication mechanisms, the security threats faced by the clients and server, and credential theft and remote attacks implemented through unregistered devices anywhere on the internet.

The remainder of this paper proceeds as follows. Section 2 establishes the dual-password login-authentication system. Section 3 provides an overview of the relevant password functionalities. Sections 4 and 5 make the unnecessary functionalities useless. Section 6 defines password policies and lists a few immediate benefits. Section 7 details the novel dual-identity authentication method. Section 8 develops system routines. Section 9 describes the login process for non-smartphone devices. Section 10 concludes the paper.

## 2 DUAL-PASSWORD LOGIN-AUTHENTICATION SYSTEM

In this work, a user-selected password is converted into a complex password to create a pair of passwords, and then this pair of passwords is integrated into a traditional user authentication mechanism to establish a dual-password login-authentication mechanism.

### 2.1 Creating a pair of login and authentication passwords

Specifically, we employ a single-character conversion technique [11] to randomly convert a set of login characters selected by a user into a sequence of strings, as shown in Figure 1, and then all the strings are combined in a special manner to generate a longer and more complex password. Thus, a pair of passwords is created.

| Login Character | Character Digit | Converted String |
|---|---|---|
| b | 6 ▼ | 3Mo&(E |
| @ | 3 ▼ | vX# |
| 0 | 5 ▼ | z%9CP |
| N | 2 ▼ | ?G |
| 8 | 3 ▼ | d$L |
| m | 1 ▼ | Q |

b@0N8m

Figure 1: The single-character conversion units

#### 2.1.1 Login password

In Figure 1, a user-selected login character "b" is randomly converted by the first single-character conversion unit into a string "3Mo&(E" according to the character digit "6" selected from the first drop-down menu of the Character Digit column. Thus, a single character is converted into a 6-digit string. The other five individual characters (i.e., "@", "0", "N", "8" and "m") are also randomly converted into five strings (i.e., "vX#", "z%9CP", "?G", "d$L" and "Q", respectively) by the other five conversion units after the corresponding digits (i.e., "3", "5", "2", "3" and "1") are sequentially selected. Thus, the set of login characters "b@0N8m" is converted into a sequence of strings (i.e., "3Mo&(E", "vX#", "z%9CP", "?G", "d$L" and "Q").

This is the first randomization process imposed on the set of login characters "b@0N8m", which can be defined as a login password and may be manually entered into the password field by the user.

#### 2.1.2 Authentication password

After further randomizing the login password, as shown in Figure 2, all the strings are sequentially shuffled together to form a longer string according to a sequence of special rules called shuffling labels. Each label in the drop-down menu of the Shuffling Label column represents a rule for inserting the string on the left side of the label into the preceding string or a temporary string to form another temporary string.

| Converted String | Shuffling Label |
|---|---|
| 3Mo&(E |  ▼ |
| vX# | 4F ▼ |
| z%9CP | 16R ▼ |
| ?G | 13F ▼ |
| d$L | 13R ▼ |
| Q | 5F ▼ |

3MovQX#&(EPC9L$d?G%z

Figure 2: Process of generating the authentication password

The labels are composed of digits and uppercase letters. For each insertion, the insertion point must first be determined; therefore, the label is prefixed with a number in front of the uppercase letter that determines the character order in the string. Each string on the left side of the label is inserted as a whole at the selected insertion point on the previous string or a temporary string.

In the first label "4F", the number "4" represents the position of the insertion point selected for the string "3Mo&(E". There are 7 insertion points from the left side of the digit "3" to the right side of the letter "E". The character order of the string is indicated by the uppercase letter "F", which represents the forward order (i.e., "vX#"), while the letter "R" represents the reverse order (i.e., "#Xv"). Therefore, the label "4F" means that the forward-order string "vX#" is inserted as a whole into the fourth insertion point on the string "3Mo&(E" to form a temporary string: "3MovX#&(E". The label "4R" means that the reverse-order string "#Xv" is inserted as a whole into the fourth point on the string "3Mo&(E" to form a different temporary string: "3Mo#Xv&(E".

According to the second label "16R", the third string is reversed to "PC9%z" and then inserted as a whole into the 16th insertion point of the first temporary string "3MovX#&(E". However, there are only 10 insertion points in the temporary string, and points 11 to 16 are invalid. Therefore, the reverse-order string "PC9%z" can only be inserted into the 10th point of the first temporary string, thus generating the "3MovX#&(EPC9%z" as the second temporary string.

The third label, "13F", is used to generate the third temporary string: "3MovX#&(EPC9?G%z". With the fourth label "13R", the fourth temporary string is generated: "3MovX#&(EPC9L$d?G%z". When the last individual character "Q" is inserted based on the label "5F", where "F" does not work on a single character, the final temporary string is generated as "3MovQX#&(EPC9L$d? G%z", which is a 20-digit character string.

The final temporary string "3MovQX#&(EPC9L$d? G%z" can be defined as an authentication password, which can be generated by converting the user-selected login password.

All the shuffling labels can be randomly selected, and the same label may even be selected repeatedly. In addition, more shuffling labels can be designed simply by changing the label digits and/or label letters.

### 2.1.3 Creating a pair of login and authentication passwords

After the two randomization processes are completed, a user-selected 6-digit login password is converted into a 20-digit complex authentication password that contains all four character classes. Therefore, a 6-20-character pair of login and authentication passwords is created after all the pair parameters (i.e., the character digits and the shuffling labels) are sequentially selected. The length of the authentication password is equal to the sum of the digits in the Character Digit column, i.e., the total number of string characters in the Converted String column.

In each single-character conversion unit, the login character or string is allowed to be modified without changing the mapping relationship between them; that is, the Character Digit column cannot be changed once selected. For example, the login character "b" in the first single-character conversion unit may be replaced with a different character (e.g., "a") without changing the string "3Mo&(E". In contrast, any character in the string can be replaced with a different character, such as replacing "o" in the string with "k", to form a new string "3Mk&(E" without changing the login character "b". The replacements can be implemented in the registration state.

In the 6-20-character password pair, both passwords are so different that the login password provides no clear sign for finding any clue about the authentication password, and vice versa. This not only means that the passwords in the pair are actually unrelated to each other but also enables their usability and security to both be achieved to some extent.

This work is intended to implement secure login usability for the client; that is, the login password should be simple and easy to remember and type without losing system security so that the user's needs and preferences can be fully satisfied and usable authentication security can be simultaneously achieved for the server; that is, the authentication password should be long and complex enough to sufficiently satisfy the security requirements of the server. To achieve these goals, the pair of passwords is integrated into a traditional user authentication mechanism to establish the dual-password login-authentication mechanism. Upon completing the integration process, the pair of passwords can be further set to be completely different to enhance the secure login usability and usable authentication security of the developed system.

### 2.2 Establishing a dual-password login-authentication mechanism

The dual-password login-authentication mechanism can be established by integrating the created pair of passwords into a traditional authentication mechanism. [15]. Two processes are included in this mechanism. One is the login process, which is defined as the process of the user entering their login password into the password field. The other is the authentication process, which is defined as the process of converting the login password into an authentication password and then authenticating the user by using the generated authentication password. In this mechanism, the login process functions to initiate the authentication process. This dual-password login-authentication mechanism is designed primarily for use on non-prepaid smartphones, while other user devices may be logged into the system via his or her registered smartphone.

Based on the integration of the system with the password pair, the login process, i.e., the process of logging the user into the system, can interact with the outside world using the login password. In contrast, the authentication process can run in the background. The purpose of having it run in an isolated environment is to prevent users from interacting with the authentication process. Based on such configurations, the isolated authentication process can be completely managed by the system. One result of this management scheme is that the pair parameters will be selected only by the system thereafter.

Accordingly, once the valid login characters are entered via the registered smartphone, they are immediately converted into a sequence of strings according to the character digits (shown in Figure 1) that are sequentially selected by the system, and then all the strings are shuffled together to generate the authentication password according to the shuffling labels (shown in Figure 2) that are sequentially selected by the system. The parameter selection steps depend on the password policies, i.e., the length and complexity requirements (Section 6). The authentication password is then compared with the registered authentication password stored in the database to authenticate the user.

Other results of the above management scheme are that the user no longer needs to memorize and enter the authentication password, thereby making it a memory- and input-free password. The authentication password is stored only on the system database; that is, it no longer needs to be stored on any local devices of the user.

The generated authentication password is a temporary password. After a successful authentication process, it is invalidated to prevent further use.

## 3 PASSWORD FUNCTIONALITY

As used in this paper, the term "local" refers to an action performed by a user via his or her smartphone that has registered with the service provider. The term "nonlocal" refers to an action performed by a party via any device other than the user's smartphone that has been registered with the service provider. For example, a local login attempt refers to an attempt by a user to enter their registered login credentials via their registered smartphone. A nonlocal login attempt refers to an attempt by a party to enter any credentials via any unregistered device.

Typically, a text string can be typed into computers, meaning that one fundamental attribute of such a string is how typable it is. If a string is defined as a password, it contains user secrets and thus can be used to verify the identity of the user, indicating that another fundamental attribute of the text password is authenticatable. Therefore, a text password must be typable and authenticatable and have typable and authenticatable functionalities.

Similar to the usability-security tradeoff encountered in single-password authentication mechanisms, the typable functionality also comes at the expense of the authenticatable functionality for the single password, in which the former is characterized by usability and the latter is described by security. Due to the presence of such opposite attributes, they are unlikely to be truly balanced within single-password mechanisms.

As noted above, any authentication mechanism involves a login process and an authentication process. Each process actually needs only one password functionality, either being typable or being authenticatable. The login process only requires a password to be typable and to have the typable functionality rather than being authenticatable. The authentication process only requires a password to be authenticatable and to have the authenticatable functionality rather than being typable.

A password is worth stealing not only because it contains user secrets and therefore is authenticatable but also because its typable functionality is indispensable for thieves to hack into user accounts using stolen passwords via unregistered devices from anywhere on the internet.

The typable and authenticatable functionalities are essential not only for any user authentication mechanism but also for credential theft and remote attacks (i.e., nonlocal login attacks). In the proposed dual-password login-authentication mechanism, the login password is defined with only the typable functionality to perform the login process, whereas its undefined authenticatable functionality is not necessary and may be discarded. In contrast, the authentication password is defined with only the authenticatable functionality to implement the authentication process, whereas its undefined typable functionality is not necessary and may be discarded. One result of such configurations is that the two necessary functionalities no longer affect each other since they belong to different passwords. Therefore, they can be further set to fully satisfy the needs and requirements of the client and the server, respectively. Another result is that the unnecessary functionalities can be either disabled or invalidated to the extent that they cannot be utilized by malicious parties, thereby rendering the corresponding passwords useless for credential theft and nonlocal login attacks.

## 4 DISABLING THE AUTHENTICATABLE FUNCTIONALITY OF THE LOGIN PASSWORD

Generally, the first step of a nonlocal login attack involves stealing a user's credentials from either the client or server since they contain the user's secrets and thus are authenticatable. In a single-password authentication mechanism, the password must be associated with identity information that is readily available to the public, such as the user's name, email address, phone number, IMEI, or SIM card ID. The use of such information as the user's identity is an inherent weakness of single-password authentication mechanisms since credential theft depends primarily on this strategy.

It is clear that the authenticatable functionality of the login password is responsible for various cyberattacks on the client. Since no one can prevent others from stealing the user's credentials, in addition to encrypting the password, disabling the authenticatable functionality of the login password is likely the only measure for preventing credential theft.

Based on the definition presented in Section 2.2, the login password is not used to authenticate the user, making it unauthenticatable. However, this unnecessary functionality is a key risk factor for client login security. With the integration of the password pair into the system, each password no longer impacts the other password, which offers an opportunity to eliminate the risk factor by disabling the authenticatable functionality. This objective can be easily achieved as long as the login password that must be entered in the password field is not associated with the identity of the user, which is related mainly to public personal information. As a result of this configuration, the login password can no longer reveal the user's secrets, rendering it useless for credential thieves.

Accordingly, the login password and the login process are no longer related to authentication events, meaning that they are immune to password-based hacking attacks against the client. Therefore, credential theft can be prevented as a result of this local login immunity, thus achieving the highest level of secure login usability for the client.

## 5 INVALIDATING THE TYPABLE FUNCTIONALITY OF THE AUTHENTICATION PASSWORD

The second step of the nonlocal login attack process is launching an attack using stolen credentials. The key to this type of attack is that it can be implemented via the unregistered devices of the thief from anywhere on the internet, which is another inherent weakness of single-password authentication mechanisms. Currently, no effective measures are available for preventing such nonlocal login attacks.

It is clear that the typable functionality of the authentication password is responsible for performing nonlocal attacks. Because no one can stop others from launching such attacks via their devices, in addition to encrypting the password, invalidating the nonlocal input or the typable functionality of the authentication password may be the only measure for preventing such attacks.

Based on the definition presented in Section 2.2, the authentication password is not used to log in, making it untypable. However, this unnecessary functionality is another key risk factor that directly threatens user accounts. Without sacrificing password functionalities, this factor can be eliminated by invalidating nonlocal inputs or the typable functionality of the authentication password. This objective can be achieved by having the authentication process run within an isolated environment and making it incapable of interacting with the outside; therefore, the authentication password can no longer be used to log in through any client device. Furthermore, when the password field is configured to only accept the login password during local login attempts (Section 6), all nonlocal inputs can be invalidated.

The results of the aforementioned configurations are that the authentication password and process are no longer related to the login events and that the nonlocal inputs or the typable functionality of the authentication password can be invalidated, even though this password can still be typed into the password field. As a result, this nonlocal input invalidation scheme can prevent remote attacks, thereby achieving the highest level of usable authentication security for the server side.

## 6 PASSWORD POLICY

In a single-password mechanism, the password field cannot be used to prevent nonlocal inputs. However, in the proposed dual-password login-authentication mechanism, the password field can be a powerful and effective means for disabling such inputs. The prerequisite for achieving this goal is that the pair of passwords must be as different in terms of length and complexity as possible. In this section, much stricter password strength (i.e., length and complexity) policies are defined so that the password field can be configured to distinguish between the passwords in the pair and thus invalidate nonlocal inputs.

### 6.1 Password length

One approach for achieving the above policies is to specify the lengths of the passwords in the pair. The password length [2, 10, 12, 19, 22] can be set as a range. In this work, the user may be permitted to determine the length of the login password within a certain range specified by the system, such as permitting passwords ranging from five to fifteen characters. Such a length specification can satisfy the needs and preferences of most users and also fulfill the requirements for users who might have special needs for longer passwords. In contrast, the authentication password may be up to twenty characters in length, which is long enough to satisfy the password length-based security requirements during the isolated authentication process.

### 6.2 Password complexity

Another approach for achieving the above policies is to specify the password complexity (i.e., the character types) [2, 12, 13, 16, 19, 21]. The login password "b@0N8m", as shown in Figure 1, might not be complex, but it contains four character classes and thus has poor usability. In this work, the login password is specified to contain only lowercase letters, digits, or a combination of the two, which are the valid characters in the dual-password mechanism. In contrast, the authentication password is specified to contain at least four character classes, and the first four characters of each authentication password must contain either one uppercase letter or one symbol.

Due to their memory- and input-free characteristics, any characters (including Unicode [2] but excluding spaces) that are capable of being processed by computers can be accepted in the composition of the authentication password. The use of such characters can help make the authentication password more complex, thereby satisfying the most stringent security requirements for the authentication process.

### 6.3 Corresponding results

The above password policies are designed from the perspectives of both the client and the server to further enhance their local login immunity and nonlocal input invalidation effects, thereby achieving secure login usability and usable authentication security, which can lead to the following advantages.

#### 6.3.1 Password field configurations

Due to the large difference between the passwords contained in the pair, the user interface can now be used to detect password strength violations, and the password field can be configured to only accept valid characters. This makes the authentication password no longer acceptable by the dual-password login-authentication system through the password field from anywhere on the internet. That is, any passwords containing invalid characters are useless in any login attempts.

### 6.3.2 User experience

We consider secure login usability to be a prerequisite for a good user experience [2, 4, 17, 18]. The failure to provide users with secure login usability leads to a poor user experience, whereas asking them to comply with various security policies significantly increases user frustration. In the user-centered password policies described herein, we reduce the burden imposed on users to the minimum level by using only lowercase letters and/or digits to compose the login password, giving them unprecedented convenience throughout the login process. In contrast, we impose almost all the strict requirements on the system to achieve usable authentication security.

The implementation of secure login usability, i.e., providing the best user experience, is crucial for the dual-password mechanism so that usable authentication security can be implemented. Otherwise, significant authentication issues arise as a result of the failure of the system to provide users with the secure convenience of entering their login passwords.

### 6.3.3 Password reuse

Password reuse [10, 16, 19, 22, 27] is considered the second source of high-success guesses [19], and the ever-increasing number of online services has accelerated the trend toward reuse. From the perspective of secure login usability, password reuse is an inherent weakness of single-password authentication mechanisms. Provided that user-selected passwords are still authenticatable and thereby must comply with complicated password policies, poor user experiences will inevitably lead users to work around such policies with some tricks.

In the proposed dual-password login-authentication mechanism, password reuse is no longer a problem. First, it is impossible for users to reuse or share their system-managed authentication passwords because they do not possess them. Second, due to the above password policies, it is useless to input the authentication password since it cannot be accepted by the password field from anywhere on the internet via any device.

In contrast, the likelihood that the login password is reused across multiple accounts may be very high, although this strategy is not recommended. However, the key to successfully logging into the dual-password login-authentication system is the use of a registered smartphone. That is, only by entering the registered login password during a local login attempt can the user log into the system. Otherwise, even the same login password will be invalidated when it is entered during nonlocal login attempts.

A user may have many registered login passwords for their accounts; some of them may even be identical. However, they can use only one universally unique smartphone to register and access their accounts. In the worst case, if a malicious party has obtained the user's credentials for a specific account in some way, the only thing that the malicious actor can do is launch a nonlocal login attack using the stolen credentials. Thus, this nonlocal input will necessarily be invalidated.

### 6.3.4 Password modification

The dual-password login-authentication system provides password modification functions. The above password policies may provide guidance for implementing such modifications. On the basis of the original pair of passwords, their lengths and/or character types can be further modified in the registration state by the user and system, respectively.

Based on the characteristics of the single-character conversion process described in Section 2.1.3, the login password can be modified by the user based only on their needs and preferences rather than complying with strict policies or requirements. If the user prefers a simpler password, they can replace the original login password "b@0N8m" with a simpler string, e.g., "abcdef" or "123456". Considering its irrelevance to authentication events, modifying the login password does not affect system security.

Due to the memory- and input-free characteristics of the authentication password, it does not need to be frequently modified. Only when the system deems it necessary will a push notification with an "Accept" button be sent to the user's smartphone for them to accept the modification request. The system will then store all the results in its database.

## 7 CREDENTIAL REGISTRATION

The proposed dual-password login-authentication system involves a pair of passwords, which makes it possible to define two identities and associate them with the pair of passwords; therefore, we have a dual-identity authentication scheme. "The more identity factors employed, the more robust the authentication system" [28]. Typically, the identity of a user is associated with their authentication password to grant the user access to the password page and later to their account at the system server.

In this paper, a novel identity is defined for the login process, and the corresponding process identifier can be associated with the login password to grant permission for a local login attempt to initiate the authentication process. As the system is designed for use on smartphones, the registration and login steps must be performed through the user's registered smartphone.

### 7.1 Registration of the identity of the user

During the process of registering an authentication password, the system requires a user to use his or her non-prepaid smartphone to submit personal information, converts the user-entered login password, and generates the authentication password. The system can also collect identification information (such as the IMEI number, the SIM card ID, or any other information that can uniquely identify the smartphone). The system server then creates and stores the submitted and collected information in its database.

Typically, the system can associate the identity of the user with their smartphone, select a unique phone number as the identifier and associate the smartphone identifier of the user with their authentication password. The collected identification information can be used to help identify whether the smartphone is the true device that is registered with the username of the created user account. Subsequently, the system can grant the user access to the password page and later to the user account at the system server after the authentication password is generated.

### 7.2 Registration of the identity of the login process

During the process of registering the login password, any information that is uniquely related to the login process can be selected as the process identifier and associated with the login password. Figure 3 is an illustration of a quasi-matrix password converter resulting from the combination of Figures 1 and 2, providing an overview of how a 6-digit login password is converted into a 20-digit authentication password. The login password "b@0N8m" in Figure 1 is modified to "abc123" based on the password policies described in Section 6.

| Login Character | Character Digit | Converted String | Shuffling Label |
|---|---|---|---|
| b | 6 | 3Mo&(E | |
| @ | 3 | vX# | 4F |
| 0 | 5 | z%9CP | 16R |
| N | 2 | ?G | 13F |
| 8 | 3 | d$L | 13R |
| m | 1 | Q | 5F |

Figure 3: The quasi-matrix password converter

During the process of creating a user account, such a converter can be established by the system to convert the user-selected login password into an authentication password. For a different account, the converter is different.

Clearly, the converter is part of the isolated authentication process, meaning that it is concealed and inaccessible to users. However, the elements in the converter are connected to the user-selected login characters, i.e., the login password. As the result of such unique connections, the elements can be unique to the process of entering the login characters.

Considering that the authentication process is completely managed by the system, the unique identifier of the login process can only be selected by the system from the converter elements and then associated with the login password. More importantly, the elements other than those in the Login Character column are unrelated to the personal information that is readily available to the public, such as the user's name, email address, phone number, IMEI number, and SIM card ID.

There are multiple ways to select the process identifier from the converter. It may not be necessary to select the whole converter as the process identifier. In fact, an element row including the login character, an element column excluding the Login Character column, or a combination of a few elements randomly selected from the converter can be unique to the login password. Thus, any one of the rows, columns, or element combinations may be selected as the process identifier that will be associated with the login password. Therefore, the system can grant local login attempt permission to initiate the authentication process.

A converter can be established during the process of registering a user account via the user's smartphone and can be reestablished only during a local login attempt by the system for generating the authentication password. The converter actually binds the specific user account to the corresponding unique smartphone via the process identifier, meaning that every established converter is legitimate. Therefore, it is not possible for a party to integrate a counterfeit converter, if any, into the isolated authentication process to bind the specific user account with the party's unregistered smartphone. Therefore, it is highly unlikely that a nonlocal login attempt can deceive the identifier verification process into initiating the authentication process.

Furthermore, the converter functions as a hash algorithm [2, 8], mapping a user-selected variable-length login password to a system-generated fixed-length authentication password. However, unlike in the traditional algorithm, the intermediate elements of the converter are available to the login password so that they can be used by the system to define a unique process identifier for the login password to be associated with. The primary characteristics of the process identity are as follows.

- The identity of the login process can be completely controlled by a trustworthy service provider with the user's explicit consent and authorization [23].
- This process identity solution discloses no amount of identifying information about the user, as it does not contain any personal information and is completely useless in other scenarios [23].
- Utilizing any computer-processable special characters in the composition of the converted strings in Figure 3 makes the process identifier much more difficult to fake or spoof.
- The identity verification process can be implemented via the interior of the system to verify the linkage between the registered smartphone and the user account through the system-determined unforgeable process identifier.

- The process identifier does not need to be transmitted through cyberspace or entered manually by users, which can effectively prevent security threats, such as identity theft and fraud.

The process identity is related to a knowledge factor [2, 24, 28], which is something the user knows. However, in the novel dual-identity authentication scheme, another difference is that the user does not need to know the details of the authentication factor since it can be completely managed by a trustworthy service provider on the user's behalf. That is, the process identity is something the user is aware of, but they do not know the details of the corresponding process identifier.

During the credential registration process, a username and a pair of passwords are registered in the database. However, the authentication password no longer needs to be stored on any device possessed by the user [25], while the login password can be managed only by the user.

The associations between the identifiers and the passwords in the pair may change if either of them is modified. The system server can then store the results and update the associations between the identifiers and the passwords.

## 8 SYSTEM ROUTINE

The dual-password system can interact with the outside world in two ways, i.e., the username and password fields at the user interfaces, which might be maliciously used to log in. To solve such potential threats, we develop system routines to detect the input fields of the user interfaces and disable any unauthorized inputs.

### 8.1 System routine of the smartphone identifier

When a user accesses the user interface via their smartphone, in response to the input provided on the username page, the dual-password login-authentication system begins with a system routine to detect whether the smartphone has registered with the entered username by checking the database for a matching smartphone identifier and the corresponding identification information associated with the user account. For a user who has already created an account with the service provider, the system routine can identify their smartphone as the one that has registered with the service provider, thereby identifying that this is a local login attempt. Whenever the user logs in, the username associated with their account may appear in the username field displayed on their smartphone, and this identified local login attempt can be granted access to the password page to attempt to initiate the authentication process.

For a party who has not yet created an account with the service provider, in response to the input, the username field displayed on their smartphone will be blank; therefore, they may have an opportunity to register an account. However, if they choose to enter a username with the intention of logging in, the system routine will be called to identify whether the smartphone has registered with the entered username by searching for a matching smartphone identifier in the database. A mismatch can indicate that this is a nonlocal login attempt; therefore, the password field displayed on the party's smartphone will be grayed out to disable further inputs. As a result of this grayed-out interface for protection, the party will have nowhere to enter their password, meaning that the identified nonlocal login attempt is invalidated.

### 8.2 System routine of the process identifier

In response to the input provided in the password field, the system routine is called again to check the database to determine whether the user-entered login password matches the process identifier. For a registered user, the system routine can identify the login password in a local login attempt by searching for a matching process identifier in the database and grant the identified local login attempt permission to initiate the authentication process. Once initiated, the authentication process reestablishes the quasi-matrix password converter to convert the user-entered login password for generating the authentication password and compares it with the registered authentication password stored in the database. If they match, this local login attempt can be granted access to the user account at the service provider.

In the worst case, we assume that an attacker might have somehow tricked the service provider into switching the user's phone number to a SIM card contained in the attacker's smartphone [14, 26] and thus gained access to the password page for entering the stolen login password. However, because the attacker's smartphone has never been bound to the legitimate user's accounts by way of registration, the system server cannot find a process identifier associated with this smartphone. Consequently, the password input will be identified as a nonlocal login attempt, and the attacker will be locked out of initiating the authentication process.

### 8.3 System routine of password strength

The system routine can also respond to the input provided in the password field by detecting password strength violations. Once a violation is detected, the identified nonlocal login attempt will be locked out of initiating the authentication process. However, for a local login attempt, the system may define a maximum number of consecutive unsuccessful login attempts, e.g., three attempts. Therefore, a predefined number of local login attempts can be granted permission to initiate the authentication process.

## 9 LOGGING NON-SMARTPHONE DEVICES INTO THE SYSTEM

The dual-password login-authentication system is designed mainly for use on non-prepaid smartphones, through which users may register and log in. However, other devices (such as desktops, laptops, and tablets) can also be logged into the system through the registered smartphone. To this end, the user interface may provide the use of technology (such as QR codes or biometric sensors) for the registered smartphone to log those devices into the system. For example, the user interface may configure a QR code for a desktop computer to be logged in, and the user's smartphone may be configured to scan the QR code using its camera. Thus, a user may use their registered smartphone to scan the QR code displayed on the screen of their computer and to directly gain access to the user account.

When a third party scans the QR code on their computer screen, they may either log into their own account if they have one with the service provider or be directed to the registration interface. Due to the legitimacy of the quasi-matrix password converter, it is not possible for them, through their smartphone, to trick the system into reestablishing the converter in association with other users' accounts. That is, the party cannot log their computer into other users' accounts by means of the QR code.

## 10 CONCLUSION

Users prefer secure login usability, which leads to the best user experiences, whereas service providers require usable authentication security, which results in more secure organizations. With the integration of a password pair into the system, striking such a balance between usability and security is no longer necessary. On the basis of these improvements, the password pair can be further set to fully satisfy the needs and requirements of the client and the server without sacrificing each other's functionalities.

Credential theft and nonlocal login attacks have always been the most serious threats to user authentication mechanisms, where the former results from the discarded authenticatable functionality, while the latter is derived from the discarded typable functionality. As this tradeoff is resolved, these once-uncontrollable password functionalities are disabled or invalidated, making the pair of passwords no longer useful for conducting nonlocal login attacks.

In the proposed dual-password login-authentication mechanism, a novel process authentication factor can be defined to verify the truly unique identity of the login process, which is characterized by concealment, inaccessibility, and independence from any public information about the user. As a result of this process identifier with such unique characteristics, the password field can be protected.

## REFERENCES

[1] Brian Jackson. 2017. Security versus usability: overcoming the security dilemma in financial services. (October 2017). Retrieved August 15, 2023 from https://www.microsoft.com/en-us/industry/blog/financial-services/2017/10/19/security-versus-usability-overcoming-the-security-dilemma-in-financial-services/

[2] Paul A. Grassi, James L. Fenton, Elaine M. Newton, Ray A. Perlner, Andrew R. Regenscheid, William E. Burr, and Justin P. Richer. 2017. Digital Identity Guidelines, *Authentication and Lifecycle Management*. NIST SP 800-63b (updates 03-02-2020). https://doi.org/10.6028/NIST.SP.800-63b

[3] Ömer Aslan, Semih Serkant Aktyğ, Merve Ozkan-Okay, Abdullah Asim Yilmaz, and Erdal Akin. 2023. A Comprehensive Review of Cyber Security Vulnerabilities, Threats, Attacks, and Solutions. *Electronics* 12, 1333 (March 2023), 1-42. https://doi.org//10.3390/electronics12061333

[4] Denis Feth. 2015. User-centric security: optimization of the security-usability trade-off. In *Proceedings of the 2015 10th Joint Meeting on Foundations of Software Engineering*, August, 2015, 1034-1037. https://doi.org/10.1145/2786805.2803195

[5] Farrukh Sahar. 2013. Tradeoffs between Usability and Security. *IACSIT International Journal of Engineering and Technology* 5, 4 (August 2013), 434-437. http://dx.doi.org/10.7763/IJET.2014.V5.591

[6] Khalid T. Al-Sarayreh, Lina A. Hasan, and Khaled Almakadmeh. 2016. A trade-off model of software requirements for balancing between security and usability issues. *International Review on Computers and Software (IRECOS)* 10, 12 (December 2015), 1157-1168. https://doi.org/10.15866/IRECOS.V10I12.8094

[7] M. Angela Sasse, Matthew Smith, Cormac Herley, Heather Lipford, and Kami Vaniea. 2016. Debunking security-usability tradeoff myths. *IEEE Security & Privacy* 14, 5 (Sept.-Oct. 2016), 33-39. https://doi.org/10.1109/MSP.2016.110

[8] Valentin Mulder, Alain Mermoud, Vincent Lenders, and Bernhard Tellenbach. 2023. *Trends in Data Protection and Encryption Technologies*. (eBook, 1st. ed.). Springer Cham, Berlin. https://doi.org/10.1007/978-3-031-33386-6

[9] Suzanne Widup, Alex Pinto, Dave Hylender, Gabriel Bassett, and Philippe Langlois. 2022. *Verizon: Data Breach Investigations Report*, Verizon Technical Report, May 2022. http://dx.doi.org/10.13140/RG.2.2.28833.89447

[10] Ponemon Institute: The 2019 State of Password and Authentication Security Behaviors Report, Phoemon Institute LLC, January 2019. Retrieved February 20, 2024 from https://www.yubico.com/press-releases/yubicos-2019-state-of-password-and-authentication-security-behaviors-report/

[11] Yun Su and Mo Xi. Password generation method which satisfies the requirement for security and usability simultaneously. Patent No. PCT/IB2019/052719, Filed April 3rd, 2019.

[12] Kristen K. Greene, John Kelsey, and Joshua M. Franklin. 2016. Measuring the Usability and Security of Permuted Passwords on Mobile Platforms. (NIST, Gaithersburg, MD), NISTIR 8040.

https://doi.org/10.6028/NIST.IR.8040

[13] Dinei Florencio and Cormac Herley. 2010. Where Do Security Policies Come From? In *Proceedings of the Sixth Symposium on Usable Privacy and Security (SOUPS '10)*, July 14-16, 2010, Remond, WA USA, 1-14. https://doi.org/10.1145/1837110.1837124

[14] Kevin Lee, Benjamin Kaiser, Jonathan Mayer, and Arvind Narayanan. 2020. An Empirical Study of Wireless Carrier Authentication for SIM Swaps. In *Proceedings of the Sixteenth Symposium on Usable Privacy and Security (SOUPS '20)*, August 9-11, 2020, Princeton, NJ, USA, 61-79. https://www.usenix.org/conference/soups2020/presentation/lee

[15] Yun Su and Mo Xi. Method for a login-authentication system using a pair of login and authentication passwords. Patent No. PCT/IB2023/061846, Filed November 23rd, 2023.

[16] Anupam Das, Joseph Bonneau, Matthew Caesar, Nikita Borisov, and XiaoFeng Wang. 2014. The Tangled Web of Password Reuse. In *Proceedings of the Network and Distributed System Security Symposium (NDSS '14)*, February 23-26, 2014, San Diego, CA, USA. http://dx.doi.org/10.14722/ndss.2014.23357

[17] Allam Hassan Allam, Ab Razak Che Hussin, and Halina Mohamed Dahlan. 2013. User experience: challenges and opportunities. *Journal of Information Systems Research and Innovation (JISRI)*, 3, (February 2013), 28-36.

[18] Panagiotis Zagouras, Christos Kalloniatis, and Stefanos Gritzalis. 2017. Managing User Experience: Usability and Security in a New Era of Software Supremacy. In *International Conference on Human Aspects of Information Security, Privacy and Trust (HAS '17)*, May 13, 2017, vol. 10292, Springer, Cham, Switzerland, 174-188. https://doi.org/10.1007/978-3-319-58460-7_12

[19] Hana Habib, Jessica Colnago, William Melicher, Blase Ur, Sean Segreti, Lujo Bauer, Nicolas Christin, and Lorrie Cranor. 2017. Password Creation in the Presence of Blacklists. In *Proceedings of the Workshop on Usable Security (USEC '17)*, Feb 26-Mar 1, 2017, San Diego, CA, USA. http://dx.doi.org/10.14722/usec.2017.23043

[20] Matt Weir, Sudhir Aggarwal, Micharl P. Collins, and Henry Stern. 2010. Testing Metrics for Password Creation Policies by Attacking Large Sets of Revealed Passwords. In *Proceedings of the 17th ACM Conference on Computer and Communications Security (CCS '10)*, October 4-8, 2010, Chicago, Illinois, USA, 162-175. http://dx.doi.org/10.1145/1866307.1866327

[21] Merve Yildirim and Ian Mackie. 2019. Encouraging users to improve password security and memorability. *Int. J. Inf. Secur* 18, 12 (April 2019), 741-759. https://doi.org/10.1007/s10207-019-00429-y

[22] Indira Mannuela, Jessy Putri, Michael, and Maria Susan Anggreainy. 2021. Level of password vulnerability. In *proceedings of the 1st International Conference on Computer Science and Artificial Intelligence (ICCSAI '21)*, October 28, 2021, Jakarta, Indonesia, 351-354. https://doi.org/10.1109/ICCSAI53272.2021.9609778

[23] Kim Cameron. 2005. The laws of Identity. Retrieved November 25, 2023 from http://www.identityblog.com/?p=354

[24] Aleksandr Ometov, Sergey Bezzateev, Niko Makitalo, Sergey Andreev, Tommi Mikkonen, and Yevgeni Koucheryaby. 2018. Multi-Factor Authentication: A Survey. *Cryptography* 2, 1 (2018), 1-31. http://dx.doi.org/10.3390/cryptography2010001

[25] Noam Ben-Asher, Niklas Kirschnick, Hanul Sieger, Joachim Meyer, Asaf Ben-Oved, and Sebastian Moller. 2011. On the need for different security methods on mobile phones. In *Proceedings of the 13th International Conference on Human Computer Interaction with Mobile Devices and Service (MobileHCI '11)*, Aug 30-Sept 2, 2011, Stockholm, Sweden, 465-473. https://doi.org/10.1145/2037373.2037442

[26] FBI Public Service Announcement. 2022. Criminal Increasing SIM Swap Schemes to Steal Millions of Dollars from US Public. Alert Number I-020822-PSA, February 8, 2022. Retrieved July 25, 2023 from https://www.ic3.gov/Media/Y2022/PSA220208

[27] Sai Sandilya Konduru and Sweta Mishra. 2023. Detection of Password Reuse and Credential Stuffing: A Server-side Approach. *IACR Cryptol. ePrint Arch*. 2023 (2023) 989. https://api.semanticscholar.org/CorpusID:259324658

[28] Paul. A. Grassi, Michael. E. Newton, and James. L. Fenton. 2017. Digital Identity Guidelines. NIST SP 800-63-3 (updates 03-02-2020). https://doi.org/10.6028/NIST.SP.800-63-3